\documentclass[seceq,preprint]{ptptex}

\usepackage{graphicx}

\makeatletter
\renewcommand{\theequation}{\thesection.\arabic{equation}}
 \@addtoreset{equation}{section}
\makeatother

\notypesetlogo                       
\preprintnumber[3cm]{
OU-HEP 441 \\
het-ph/0304283 \\
May 9, 2003}

\title{
New Derivation of Anomaly-Mediated Gaugino Mass in the Higher Derivative Regularization Method %
\vspace{1cm}
}

\author{
Koske \textsc{Nishihara}
\footnote{e-mail:koske@het.phys.sci.osaka-u.ac.jp} 
and Takahiro \textsc{Kubota}
\footnote{e-mail:kubota@het.phys.sci.osaka-u.ac.jp}%
}

\inst{
Graduate School of Science, Osaka University, Toyonaka, Osaka 560-0043, Japan
}

\abst{
The technique of higher derivative regularization is applied 
to the Super-Weyl-K{\" a}hler anomaly in supergravity 
coupled with chiral matters and gauge multiplet. 
Our method makes it possible to derive directly and 
diagrammatically  the  formulae for the gaugino mass 
(in the Abelian case).
 Our procedure is applied to derivation of not only the mass formulae 
known in the anomaly mediation scenario, but also 
additional terms that depend on the K{\" a}hler potential.
}

\begin{document}

\maketitle

\section{Introduction}\label{intro}

One of the most important subjects in the present-day 
 particle physics is to look for experimental signatures of 
supersymmetry (SUSY) and search for favorable  
sources of SUSY breaking mechanism . Among several others, 
extensive studies have been made to look for scenarios
 of SUSY breaking triggered by 
gravity \cite{nilles} and gauge interactions
 \cite{giudice}. Several years ago, a novel type 
of SUSY breaking mechanism was proposed  which is often 
referred to as anomaly mediation mechanism. 
It has been argued in 
Refs.~\citen{murayama} and \citen{rs}  
that the SUSY breaking is communicated through 
the super-Weyl anomaly via supergravity.

The most appealing aspects of the anomaly mediation 
is its unique predictability of soft breaking terms. For
example gaugino masses, scalar masses  and triple coupling 
of matter multiplets are given respectively by
\begin{eqnarray}
m_{\lambda}& =& \frac{1}{3} \beta_{g}  M^{*},    
\label{eq:gauginomass}
\\
m^{2}_{i} & = & -\frac{|M|}{36}^2   \!\! \left( 
\frac{\partial \gamma_i}{\partial g}\beta_{g}
          +\frac{\partial \gamma_i}{\partial y}\beta_y  \!\!\right), 
\\
h^{ijk}  & = &  \frac{1}{6} (\gamma_{i}+\gamma_{j}+\gamma_{k})y^{ijk} M^* .
\label{eq:yukawa}
\end{eqnarray}
Here indices $i$, $j$ and $k$ label the chiral 
matter multiplets.
These parameters are all determined in terms of 
the  beta functions of the gauge coupling $\beta _{g}$ 
and the Yukawa coupling $\beta _{y}$ and the anomalous 
dimensions $\gamma _{i}$ of the $i-$th chiral 
multiplet  field.  $M$ is the auxiliary field of 
the gravity multiplet whose vacuum expectation value  
is the source of SUSY breaking.

In spite of its high predictability, the anomaly mediation 
scenario entails drawbacks of its own. Namely, scalar 
partners of leptons are given  negative mass squared. 
Various ideas have been proposed to solve this tachyonic 
slepton mass problem \cite{rs}\tocite{luty}. 
They are, however, rather sophisticated and spoil the simplicity of the original proposal.

We would also like to note that there could exist  
additional terms in the formula (\ref{eq:gauginomass})-(\ref{eq:yukawa}) 
which depends on the matter K{\" a}hler potential 
\cite{moroi}. The minimal supergravity coupled with 
matter and gauge multiplets is invariant under 
the super-Weyl-K{\" a}hler transformation on the 
classical level. The quantum anomaly associated with 
this symmetry contains the K{\" a}hler potential  
on the order of $\kappa ^{2}=8\pi G_{N}$. 
Such additional terms could  hopefully change  
the nature of the slepton tachyonic problem. 

Bearing these considerations in  mind, we now examine diagrammatically  
the gaugino mass formula due to the super-Weyl-K{\" a}hler anomaly. 
We will see that the most 
suitable way to derive the gaugino mass directly 
is the use of the higher derivative  regularization method. 
It has been known for some time that this 
method alone is not always 
very helpful at one-loop 
\cite{waar,kap}. 
It becomes useful only when 
we make a combined use of other regularizations, typically 
such as Pauli-Villars (PV) method. Admitting such shortcoming, we would still like to  show that 
the higher derivative regularization combined with 
the PV method  turns out to be instrumental to uncover 
diagrammatical structures of the super-Weyl-K{\" a}hler 
anomaly and to produce formulae known in anomaly mediation.

The present paper is organized as follows.
First we recapitulate the super-Weyl-K{\" a}hler anomaly 
in $\S$ \ref{anomaly} and the higher derivative regularization 
in $\S$ \ref{higher}. We reproduce in $\S$ \ref{gaugino} 
the mass formula (\ref{eq:gauginomass}) for the case of the Abelian gauge 
group. (To avoid unessential complications, we consider only the 
Abelian case throughout.) In $\S$ \ref{kahler} our method is applied further 
to derive the terms to be added to (\ref{eq:gauginomass}),  
which depends on the K{\" a}hler potential.  
$\S$ \ref{conclusion} is devoted to conclusions.


\section{Super-Weyl-K{\" a}hler Anomaly}
\label{anomaly}

As we mentioned in  $\S$ \ref{intro}, the minimal supergravity 
coupled with  matter and gauge fields is invariant under 
the simultaneous transformation of super-Weyl and K{\" a}hler 
transformations on the classical level. The anomaly on the 
quantum level of this symmetry has been discussed in literatures to 
a considerable extent 
\cite{kap,ovrut,ferrara}.  

The fermionic contribution due to the super-Weyl-K{\" a}hler 
anomaly to the effective action has been evaluated as 
\begin{eqnarray}
\frac{g^{2}}{96\pi ^{2}}b_{0}
{\rm Tr}F_{mn}{\tilde F}^{mn}\frac{1}{\Box}
\partial _{\ell  }c^{\ell }.
\end{eqnarray}
Here the connection $c^{\ell }=b^{\ell }+4i\kappa ^{2}a^{\ell}$
consists of the auxiliary field $b^{\ell }$ in the gravitational 
superfield and the K{\" a}hler connection $a^{\ell }$. 
The first coefficient of the beta-function is denoted by $b_{0}=3T(G)-T(R)$ .
This anomaly would  be lifted to the unique 
supersymmetric expression
\begin{eqnarray}
\frac{g^{2}}{256 \pi ^{2}}\int d^{4}\Theta 2{\cal E} 
W^{\alpha }W_{\alpha }
\frac{1}{\Box }\left (
{\bar {\cal D}}^{2}-8R \right ) 
\left \{ 4b_{0}R^{\dag} +\frac{\kappa ^{2}}{3}T_{R}{\cal D}^{2} K
+  \cdots \right \}
 +{\rm h.c.}, 
\label{eq:co}
\end{eqnarray}
if supersymmetry could be  respected throughout. 
Here $K$ is the K{\" a}hler potential and 
$R$ is the gravity superfield expanded as 
\begin{eqnarray}
R=-\frac{1}{6}\left \{
M+\cdots + \Theta ^{2}\left (-\frac{1}{2}{\cal R}
-i{e_{a}}^{m}{\cal D}_{m}b^{a}
+\cdots \right )
\right \}.
\label{eq:M}
\end{eqnarray}
The Einstein scalar curvature is denoted by ${\cal R}$. 
 (The ellipses in (\ref{eq:co}) 
correspond to the sigma-model anomaly, into which 
we do not delve throughout the present paper.~)

It is easy to see that if the the auxiliary field  
$M$ in the superfield $R$ in (\ref{eq:co}) 
acquires  a vacuum expectation value, the formula 
 (\ref{eq:co}) gives a 
mass term for the gaugino in accordance with the 
formula (\ref{eq:gauginomass}). A natural question arises, 
however: Although nothing is wrong with 
the formula (\ref{eq:co}), one would wonder if we could 
derive the gaugino mass formula in a more direct diagrammatic 
way which could 
get a more insight into the regularization. It is somewhat 
puzzling that a naive look at Feynman rules 
does not provide us with  Feynman diagrams 
producing such a gaugino mass term. This puzzling situation 
is cleared only when we give details of the regularization. 
The present work is an outcome of our efforts to 
visualize the derivation of the gaugino mass in 
a more familiar way. In the following sections, 
we are going to set up and apply the method of 
higher derivative regularization combined with the 
PV-type method.  This allows us to derive the gaugino mass 
in a direct way in accordance with the formula   
 (\ref{eq:gauginomass}). (See also Ref.~\citen{boyda} for 
a diagrammatic approach to the SUSY breaking.)

\section{Higher Derivative Regularization}
\label{higher}

Here we give a general consideration of the supersymmetric 
generalization of the higher derivative regularization. 
We confine ourselves to the case of rigid supersymmetry and 
postpone the description of the supergravity case  
in $\S$\ref{gaugino} and $\S$ \ref{kahler}.   

In conventional field theories, the higher derivative regularization method 
 tells us to  modify the Lagrangian, for example,  of  the scalar field 
 $A$, in the following way \cite{fad}: 
\begin{eqnarray}
A^{\dag } D ^{2}A \longrightarrow  A^{\dag } 
D ^{2}  A - \frac{1}{\Lambda^2}
A^{\dag }(D ^{2})^{2} A. 
\label{eq:scalarcase}
\end{eqnarray}
Here $D^{2}=D_{n}D^{n}$and $D_{n}$ is the 
gauge covariant derivative.
If we take the limit $\Lambda \longrightarrow \infty$, (\ref{eq:scalarcase}) reduces to the conventional kinetic 
term of the scalar field. The supersymmetric extension of (\ref{eq:scalarcase}) 
is straightforward, and the Lagrangian of a 
superfield $Q$ is modified as Ref.~\citen{kap}
\begin{eqnarray}
{\cal L}&=& \int d^{2}\theta d^{2}\bar \theta 
\left (  Q^{\dag} e^{2gV} Q
- \frac{1}{16\Lambda ^{2}} Q^{\dag} e^{2gV}
{\bar {\cal D}}^{2} e^{-2gV}{\cal D}^{2} e^{2gV} Q \right )
\nonumber \\
& &+\left ( \int d^{2}\theta \frac{1}{2}mQ^{2}+{\rm h.c.}
\right ).
\label{eq:supercase}
\end{eqnarray}
Here $V$ is a vector multiplet. 
One can see  that (\ref{eq:supercase}) is gauge invariant 
for both Abelian and non-Abelian cases. (Here and hereafter we 
follow the convention  of  Ref.~\citen{wessbagger}.
As for the higher derivative method in SUSY case we refer the reader to Ref.~\citen{Stepanyantz:2003zb} which contains many useful formulas.)

By decomposing (\ref{eq:supercase}) into component 
fields, 
\begin{eqnarray}
Q=A+\sqrt{2}\theta \chi + F, 
\label{eq:qcomponent}
\end{eqnarray}
we can easily derive propagators of scalar($A$),  
spinor($\chi $) and auxiliary ($F$) fields. We just 
list  them up in order:
\begin{eqnarray}
& &<T(A(x)A^{\dag}(y))>=\frac{i}{\Box -m^{2}-
\Box ^{2}/\Lambda ^{2}}\delta ^{4}(x-y), 
\\
& &<T(A(x)F(y))>=<T(A^{\dag }(x)F^{\dag}(y))>
=\frac{-im}{\Box -m^{2} - \Box ^{2}/\Lambda ^{2}}
\delta ^{4}(x-y), 
\\
& &<T(F(x)F^{\dag }(y))>=\frac{i\Box }{\Box -m^{2} - 
\Box ^{2}/\Lambda ^{2}}\delta ^{4}(x-y),
\\
& & 
<T(\chi _{\alpha}(x)\chi ^{\beta}(y))>=i{\delta _{\alpha }}
^{\beta}\frac{m}{\Box -m^{2} - \Box ^{2}/\Lambda ^{2}}
\delta ^{4}(x-y),
\\
& &
<T(\bar \chi ^{\dot \alpha}(x)\bar 
\chi _{\dot \beta}(y))>=i{\delta ^{\dot \alpha }}
_{\dot \beta}\frac{m}{\Box -m^{2} - \Box ^{2}/\Lambda ^{2}}
\delta ^{4}(x-y),
\\
& &<T( \chi _{\alpha}(x)\bar \chi _{\dot \beta}(y))>=
{{\sigma }^{m}} _{\alpha {\dot \beta}} \partial _{m}
\frac{1}{\Box -m^{2} - \Box ^{2}/\Lambda ^{2}}
\delta ^{4}(x-y).
\\
\end{eqnarray}
It is easy to confirm that these propagators reduce to the 
usual ones if we take the limit $\Lambda \longrightarrow \infty$.


\section{The Gaugino Mass Formula (\ref{eq:gauginomass}) Revisited}
\label{gaugino}

Now let us extend the higher derivative method to the 
supergravity case and launch the analysis of the 
super-Weyl-K{\" a}hler  anomaly. 
We separate our calculation into two parts: 
First we consider the anomaly term corresponding to 
$4b_{0}R^{\dag}$   in the curly brackets of 
(\ref{eq:co}), and then proceed in $\S$ \ref{kahler} 
to  the next term of order 
$\kappa ^{2}$  in (\ref{eq:co}). 

\begin{figure}[b]
\begin{center}
\includegraphics*[scale=1.0]{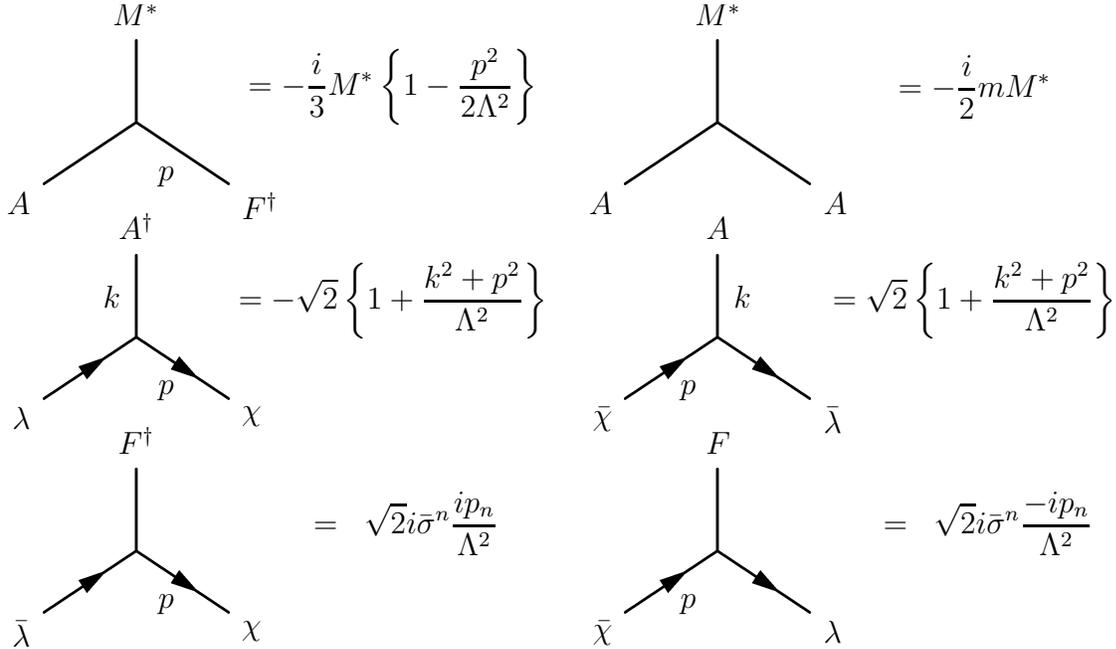}
\end{center}
\caption{
Feynman rules for vertices containing either the auxiliary field $M^{*}$ 
or matter field. Those of the PV-fields 
are the same and are omitted here.
The gaugino field is denoted by $\lambda$.}
\label{rule1}
\end{figure}

We are interested in the  evaluation of the gaugino 
mass by using the following Lagrangian 
\begin{eqnarray}
{\cal L}={\cal L}_{\rm matter}+{\cal L}_{{\rm PV}},
\label{eq:lagrangian}
\end{eqnarray}
which consists of a massless ($m=0$) chiral multiplet $Q$ in  
 ${\cal L}_{{\rm matter}}$ and of   massive  PV regulator  
 chiral field $\Phi $ (with mass $m'$) in 
${\cal L}_{{\rm PV}}$. More explicitly each term in 
(\ref{eq:lagrangian}) is expressed as 
\begin{eqnarray}
{\cal L}_{\rm matter}  &=& \!\int \!\! d^2 \Theta\  
 2{\cal E}
 \Bigg [-  
 \frac{1}{8}({\bar{\cal D}}^{2}-8R)
 \Bigg \{ Q^{\dag}e^{2gV}  Q  \rule{0mm}{6mm}
\nonumber \\
& &   
- \frac{1}{16\Lambda^2} Q^{\dag} e^{2gV}
 ({\bar {\cal D}}^2 - 8R)e^{-2gV}{\cal D}^{2} e^{2gV}
 Q   \Bigg \} \Bigg ] 
  +{\rm h.c.}, 
\label{eq:lmatter}
\\
{\cal L}_{\rm PV}  &=& \!\int \!\! d^2 \Theta\  
 2{\cal E}
 \Bigg [-  
 \frac{1}{8}({\bar{\cal D}}^{2}-8R)
 \Bigg \{ \Phi ^{\dag}e^{2gV}  \Phi  \rule{0mm}{6mm}
\nonumber \\
& &   
- \frac{1}{16\Lambda^2} \Phi ^{\dag} e^{2gV}
 ({\bar {\cal D}}^2 - 8R) e^{-2gV}{\cal D}^{2} e^{2gV}
 \Phi    \Bigg \} +\frac{m'}{2}\Phi ^{2} \Bigg ] 
  +{\rm h.c.}. 
\label{eq:lpv}
\end{eqnarray}
This Lagrangian gives many additional terms containing 
$1/\Lambda ^{2}$, and our text would be too much 
cluttered if we would write down all the Feynman rules. 
Since we are interested in the gaugino mass and therefore 
in the single insertion of the auxiliary field $M$ 
in (\ref{eq:M}),   
we pay our attention to  the  vertex containing $M$ singly. 
Some of the Feynman rules relevant to our calculation are 
illustrated in Fig. \ref{rule1}, where those containing
 the matter field $Q$ are given: the Feynman rules 
of PV-field $\Phi $ are exactly the same. 

\begin{figure}[htb]
\parbox{8.5cm}{
{\footnotesize 
\begin{description}
\item{Table 1.}
Various combinations of the propagators in~Fig.~\ref{graph1}
\end{description} 
}
\begin{tabular}{c|c c c } 
\hline \hline
 & (a) & (b) & (c) 
\\
\hline 
massless matter & $<F^{\dag}F>$ & $<\bar \chi \chi>$ & $<A^{\dag}A>$ 
\\
\hline
PV (1)& $<F'^{\dag}F'>$ & $<\bar \chi ' \chi '>$ & $<A'^{\dag}A'>$ 
\\
\hline
PV (2)& $<F'^{\dag}A'^{\dag}>$ & $<\chi ' \chi '>$ & $<A'^{\dag}A'>$
\\
\hline
PV (3)& $<A'A'^{\dag}>$ & $<\chi ' \chi '>$ & $<A'^{\dag}A'>$
\\
\hline
PV (4)& $<F'^{\dag}A'^{\dag}>$ & $< \chi ' \bar \chi '>$ & $<F'A'>$
\\
\hline
PV (5)& $<F'^{\dag}F'>$ & $<\bar \chi ' \bar \chi '>$ & $<F'A'>$
\\
\hline
PV (6)& $<A'F'>$ & $<\chi ' \bar \chi '>$ & $<A'^{\dag}A'>$
\\
\hline
PV (7)& $<A'F'>$ & $<\bar \chi ' \bar \chi '>$ & $<F'A'>$
\\
\hline 
\end{tabular}
}
  \hfill
\parbox{6cm}{
\begin{center}
\includegraphics*[scale=1.00]{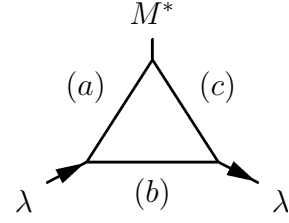}
\end{center}
\caption{
The Feynman diagram giving rise to the gaugino 
($\lambda $) mass. The combination of the 
propagators (a), (b) and (c) are given in Table 1.
\label{graph1}}
}
\end{figure}

Being equipped with the Feynman rules, we are now 
in a position to compute the triangle diagrams of the 
type  Fig. \ref{graph1} giving rise to the gaugino mass term. 
There are several combinations of propagators running 
through the triangle and they are listed in Table 1.
There are seven diagrams in which massive PV-field are circulating. Its component fields are all primed, i.e., 
\begin{eqnarray}
\Phi =A'+\sqrt{2}\theta \chi ' + F'.
\label{eq:phicomponent}
\end{eqnarray}

\noindent
The contribution of the massless chiral matter $Q$, 
the first entry in Table 1, is divergent, and gives 
an  effective action 
\begin{eqnarray}
{\cal L}_{{\rm massless}}
&\sim & 
\frac{g^{2}}{16\pi ^{2}} \frac{M^{*}}{3} \lambda \lambda \left \{
-\frac{1}{2}+2{\rm log}\left (\frac{\Lambda _{c}^{2}}{
\Lambda ^{2}}\right )\right \},
\end{eqnarray}
where $\Lambda _{c}$ is the ultra-violet cutoff. The gaugino field is denoted by $\lambda$.
The other 
contributions, PV(1)-(3),  to the effective action are given, 
respectively, by
\begin{eqnarray}
{\cal L}_{PV(1)}&=&
-\frac{2g^{2}M^{*}}{3i}\lambda \lambda \int \frac{d^{4}p}{(2\pi )^{4}}
\frac{1}{(-p^{2}-m'^{2}-(p^{2})^{2}/\Lambda ^{2})^{3}}
\frac{(p^{2})^{2}}{\Lambda ^{2}}\left ( 1-\frac{p^{2}}{
2\Lambda ^{2}}
\right )\left (1+\frac{2p^{2}}{\Lambda ^{2}}\right )
\nonumber \\
&\sim & 
\frac{g^{2}}{16\pi ^{2}} \frac{M^{*}}{3} \lambda \lambda \left \{
-\frac{1}{2}  +2 {\rm log}\left (\frac{\Lambda _{c}^{2}}{\Lambda ^{2}}
\right )+
{\cal O}(m'^{2}/\Lambda ^{2})\right \}, 
\label{eq:pv1}
\\
{\cal L}_{PV(2)}&=&
\frac{2g^{2}M^{*}}{3i}m'^{2}
\lambda  \lambda \int \frac{d^{4}p}{(2\pi )^{4}}
\frac{1}{(-p^{2}-m'^{2}-(p^{2})^{2}/\Lambda ^{2})^{3}}
\left ( 1-\frac{p^{2}}{2\Lambda ^{2}}\right )
\left (1+\frac{2p^{2}}{\Lambda ^{2}}\right )^{2}
\nonumber \\
&\sim & 
\frac{g^{2}}{16\pi ^{2}} \frac{M^{*}}{3} \lambda \lambda \left \{
-2+
{\cal O}(m'^{2}/\Lambda ^{2})\right \},
\label{eq:pv3}
\\
{\cal L}_{PV(3)}&=&
2ig^{2}M^{*}m'^{2}
\lambda  \lambda \int \frac{d^{4}p}{(2\pi )^{4}}
\frac{1}{(-p^{2}-m'^{2}-(p^{2})^{2}/\Lambda ^{2})^{3}}
\left (1+\frac{2p^{2}}{\Lambda ^{2}}\right )^{2}
\nonumber \\
&\sim & 
\frac{g^{2}}{16\pi ^{2}} \frac{M^{*}}{3} \lambda \lambda \left \{
3+
{\cal O}({m'}^{2}/\Lambda ^{2})\right \}.
\label{eq:pv6}
\end{eqnarray}
Note that (\ref{eq:pv1}) is also divergent. All the remaining 
terms become vanishing when we take the limit $\Lambda \longrightarrow \infty $, i.e., 
\begin{eqnarray}
{\cal L}_{PV(4)}=
{\cal L}_{PV(5)}=
{\cal L}_{PV(6)}=
{\cal L}_{PV(7)}=
{\cal O}(m'^{2}/\Lambda ^{2}). 
\end{eqnarray}

Although we have been dealing with a single PV-field, we can 
easily generalize our argument for the case of several 
PV-fields. In such a case 
we just superpose the sum of PV-contribution,i.e.,  (\ref{eq:pv1}),(\ref{eq:pv3}) and (\ref{eq:pv6}) 
in the following way:
\begin{eqnarray}
\frac{g^{2}}{16\pi ^{2}} \frac{M^{*}}{3}\left[ 
\left\{ 
-\frac{1}{2} + 2{\rm log}\left (
\frac{\Lambda _{c}^{2}}{\Lambda ^{2}}\right )
\right\} 
+
\sum _{i} C_{i} \left\{ 
\left (-\frac{1}{2}-2+3\right )
+2{\rm log}\left (\frac{\Lambda _{c}^{2}}{\Lambda ^{2}}\right )
\right\}
\right].
\end{eqnarray}
Here $C_{i}$ is the weight factor of each PV-fields, the 
index $i$  numbering the PV-fields.
It is now obvious that the cancellation of the ultraviolet 
divergences between the massless chiral matter  and 
PV-fields is fulfilled by 
\begin{eqnarray}
1+\sum _{i}C_{i}=0.
\end{eqnarray}
Thus the generated gaugino mass is found to be 
\begin{eqnarray}
m_{\lambda }=
\frac{g^{2}}{16\pi ^{2}} \frac{M^{*}}{3}\left \{
-\frac{1}{2}
-\left (-\frac{1}{2}-2+3\right )
\right \}
=
-\frac{g^{2}}{16\pi ^{2}}\times \frac{M^{*}}{3},
\end{eqnarray}
which agrees with previous results for $U(1)$ case. 

\section{K{\" a}hler Potential and the Gaugino Mass}
\label{kahler}

We now move on to the anomalous term  associated with the 
K{\" a}hler connection, the second term in (\ref{eq:co}). 
This term appears   on the order of 
${\cal O}(\kappa ^{2})$. For the purpose of 
evaluating it, we have to 
extend the higher derivative regularization  by 
including those of ${\cal O}(\kappa ^{2})$.
To explain the higher derivative terms of 
 ${\cal O}(\kappa ^{2})$ with 
reference to those of $\S$ \ref{gaugino},  we restart 
from the  K{\" a}hler potential of the following 
 form:
\begin{eqnarray}
K=Q^{\dag}Q+\Phi ^{\dag}\Phi ,
\label{eq:kahler1}
\end{eqnarray}
where the higher derivative terms are not yet included.
The massless chiral matter field is again denoted 
by $Q$ and $\Phi $ is a massive generic PV-chiral field. 
The Lagrangian  may be expanded in the power 
series of the gravitational constant $\kappa ^{2}$  
\begin{eqnarray}
{\cal L}&=&\frac{1}{\kappa ^{2}}\int d^{2}\Theta 2{\cal E}
\left \{
\frac{3}{8}\left ({\bar {\cal D}}^{2}-8R \right )
e^{-\kappa ^{2}K/3}
\right \}+{\rm h.c.}
\nonumber \\
&=&-\frac{6}{\kappa ^{2}}\int d^{2}\Theta 
{\cal E}R +{\cal L}_{1}+{\cal L}_{2}+{\cal O}(k^{4})
+ {\rm h.c.}, 
\end{eqnarray}
where
\begin{eqnarray}
{\cal L}_{1}&=&\int d^{2}\Theta 2{\cal E}\left \{
-\frac{1}{8}\left ({\bar {\cal D}}^{2}-8R \right )K
\right \}, 
\label{eq:L1}
\\
{\cal L}_{2}&=&\kappa ^{2}\int d^{2}\Theta 2{\cal E}\left \{
\frac{1}{48}\left ({\bar {\cal D}}^{2}-8R \right )K^{2}
\right \}.
\label{eq:L2}
\end{eqnarray}

\begin{figure}[htbp]
\begin{center}
\includegraphics*[scale=1.0]{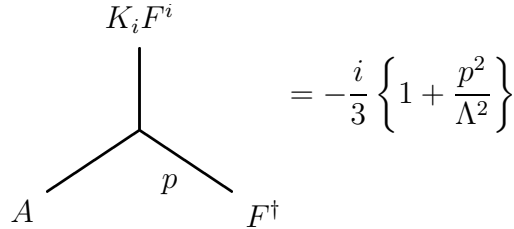}
\end{center}
\caption{
The Feynman rule for the vertex containing $K_{i}F^{i}$
}
\label{rule2}
\end{figure}

Comparing (\ref{eq:lmatter}), (\ref{eq:lpv}) 
 and (\ref{eq:L1}), we immediately 
realize that inclusion of the higher derivative terms,   
in the absence of the gauge interaction,  is achieved 
by replacing the K{\" a}hler potential (\ref{eq:kahler1}) by 
\begin{eqnarray}
K=Q^{\dag}\left \{
1-\frac{1}{16\Lambda ^{2}}
\left ({\bar {\cal D}}^{2}-8R\right ){\cal D}^{2}
\right \}Q
+\Phi ^{\dag}\left \{
1-\frac{1}{16\Lambda ^{2}}
\left ({\bar {\cal D}}^{2}-8R\right ){\cal D}^{2}
\right \}\Phi .
\label{eq:newkahler} 
\end{eqnarray}
At the ${\cal O}$($\kappa ^{2}$) level, our higher 
derivative terms are obtained by putting 
(\ref{eq:newkahler}) into (\ref{eq:L2}). 
Among many terms, those relevant to our anomaly calculation 
are
\begin{eqnarray}
{\cal L}_{2}&=&
-\frac{\kappa ^{2}}{6}\left \{
K_{i}\left ( 1 - \frac{\Box }{\Lambda ^{2}}\right )A^{i}
\right \}\left \{
F^{\dag}\left ( 1 - \frac{\Box }{\Lambda ^{2}}\right )F
+
{F'}^{\dag}\left ( 1 - \frac{\Box }{\Lambda ^{2}}\right )F'
\right \}
\nonumber 
\\
& & 
-\frac{\kappa ^{2}}{6} \left \{ 
K_{i}\left ( 1 - \frac{\Box }{
\Lambda ^{2}}\right )F^{i}\right \} \left \{
F^{ j \dag}\left ( 1 - \frac{\Box }{\Lambda ^{2}}\right )
K_{j}^{\dag}\right \} + {\rm h.c.} + \cdots .
\end{eqnarray}
Here our notations are 
\begin{eqnarray}
K_{i}\left ( 1 - \frac{\Box }{\Lambda ^{2}}\right )A^{i}
&=&A^{\dag}\left ( 1 - \frac{\Box }{\Lambda ^{2}}\right )A
+
{A'}^{\dag}\left ( 1 - \frac{\Box }{\Lambda ^{2}}\right )A' , 
\\
K_{i}\left ( 1 - \frac{\Box }{\Lambda ^{2}}\right )F^{i}&=& 
A^{\dag}\left ( 1 - \frac{\Box }{\Lambda ^{2}}\right )F
+
{A'}^{\dag}\left ( 1 - \frac{\Box }{\Lambda ^{2}}\right )F' , 
\\
F^{j \dag}\left ( 1 - \frac{\Box }{\Lambda ^{2}}\right )K_{j}^{\dag}&=& 
F^{\dag}\left ( 1 - \frac{\Box }{\Lambda ^{2}}\right )A
+
{F'}^{\dag}\left ( 1 - \frac{\Box }{\Lambda ^{2}}\right )A' .
\end{eqnarray}
Thus we arrive at the  Feynman rule depicted in 
Fig. \ref{rule2} for the vertex that contains $K_{i}F^{i}$.
Note that the vertex in Fig. \ref{rule2} is what we are interested in 
to derive the second term in (\ref{eq:co}),i.e.,
it is the $\theta$-independent term in ${\cal D}^{2}K$
\begin{eqnarray}
{\cal D}^{2}K=-4K_{i}F^{i}+\cdots .
\end{eqnarray}

\begin{figure}[t]
\parbox{8.5cm}{
{\footnotesize 
\begin{description}
\item{Table 2.}
 Various combinations of the Feynman diagrams in~Fig.~\ref{graph2}
\end{description} 
}
\begin{tabular}{c|c c c }
\hline 
\hline
 & (a) & (b) & (c)
\\
\hline
massless & $<AA^{\dag}>$ & $<\chi \bar \chi>$ & $<FF^{\dag}>$
\\
PV(i) & $<A'A'^{\dag}>$ & $<\chi ' \bar \chi '>$ & $<F'F'^{\dag}>$
\\
PV(ii) & $<A'A'^{\dag}>$ & $<\chi '  \chi '>$ & $<A'^{\dag}F'^{\dag}>$
\\
PV(iii) & $<A'F'^{\dag}>$ & $<\bar \chi '  \chi '>$ & $<A'^{\dag}F'^{\dag}>$
\\
PV(iv) & $<A'F'>$ & $<\bar \chi '  \bar \chi '>$ & $<F'F'>$
\\
\hline
\end{tabular}
}
  \hfill
\parbox{6cm}{
\begin{center}
\includegraphics*[scale=1.00]{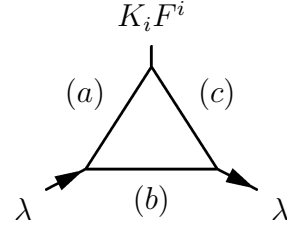}
\end{center}
\caption{
The triangle diagram producing 
$(K_{i}F^{i})\lambda \lambda $ term. Various combinations of 
the propagators (a), (b) and (c) are given in Table 2. 
\label{graph2}}
}
\end{figure}

Now the anomalous $(K_{i}F^{i})$$\lambda \lambda $ vertex 
is produced by the Feynman diagrams  in Fig. \ref{graph2}.
Various combinations of the propagators in Fig. \ref{graph2}
are summarized in Table 2.
From the Feynman rule in Fig. \ref{rule2}, it is almost obvious that 
we can borrow the loop calculations in $\S$ \ref{gaugino} 
simply by replacing 
\begin{eqnarray}
M^{*}\longrightarrow \kappa ^{2} K_{i}F^{i}
\end{eqnarray}

\noindent
To sum up, our loop calculations of Fig. \ref{graph2} are 
as follows:
\begin{eqnarray}
{\cal L}_{{\rm masless}}
&\sim & 
\frac{g^{2}}{12\pi ^{2}} \kappa ^{2} 
(K_{i}F^{i})\lambda  \lambda \left \{
\frac{1}{2}-{\rm log}\left (\frac{\Lambda _{c}^{2}}{
\Lambda ^{2}}\right )\right \},
\\
{\cal L}_{\rm PV(i)}&\sim & 
\frac{g^{2}}{12\pi ^{2}} \kappa ^{2} 
(K_{i}F^{i}) \lambda \lambda \left \{
\frac{1}{2}-{\rm log}\left (\frac{\Lambda _{c}^{2}}{
\Lambda ^{2}}\right )+
{\cal O}(m'^{2}/\Lambda ^{2})\right \},
\\
{\cal L}_{\rm PV(ii)}&\sim & 
\frac{g^{2}}{12\pi ^{2}} \kappa ^{2} (K_{i}F^{i}) \lambda 
\lambda \left \{
-\frac{1}{2}+
{\cal O}(m'^{2}/\Lambda ^{2})\right \},
\\
{\cal L}_{\rm PV(iii)}&=&{\cal L}_{\rm PV(iv)}={\cal O}
(m'^{2}/\Lambda ^{2}).
\end{eqnarray}
Again we are able to increase the number of PV-regulator 
arbitrarily, and after summing up all regulators with the weight $C_{i}$, the coefficient of the gaugino mass term 
becomes
\begin{eqnarray}
\frac{g^{2}}{12\pi ^{2}}\kappa ^{2} (K_{i}F^{i})\left[
\left\{
\frac{1}{2}-{\rm log}\left (\frac{\Lambda _{c}^{2}}{
\Lambda ^{2}}\right )\right\}
+
\sum _{i}C_{i}\left\{
\frac{1}{2}-{\rm log}\left (\frac{\Lambda _{c}^{2}}{\Lambda ^{2}}\right )-\frac{1}{2}\right\}
\right].
\end{eqnarray}
The condition canceling the ultra-violet divergence 
is the same as before
\begin{eqnarray}
1+\sum _{i}C_{i}=0,
\end{eqnarray}
and we get the gaugino mass coming from the $K_{i}F^{i}$
as 
\begin{eqnarray}
\frac{g^{2}}{16 \pi ^{2}}\times \frac{2}{3}\kappa ^{2}(K_{i}F^{i}).
\end{eqnarray}
This result is in agreement with the naive one obtained by 
using (\ref{eq:co}).

\section{Conclusion}
\label{conclusion}

In the present paper, we have examined the calculational 
basis of the super-Weyl-K{\" a}hler anomaly. We have 
pointed out that the higher derivative regularization method 
combined with  the type of PV's is useful to derive
the gaugino mass formula in a direct and diagrammatic method.
The formulae derived  in the anomaly mediation scenario are thus 
put on as familiar  footing as our old-day anomaly 
calculation.

There remain, however, several important problems. The most 
imminent is a generalization to the non-Abelian gauge group.  
In principle there does not exist a serious stumbling block 
in such a generalization, but the higher derivative 
counter terms become considerably involved. We will come to 
the non-Abelian generalization in our future publication \cite{nishihara}. 

Another problem is to reconsider the slepton mass problem, 
considering the additional contributions of order 
$\kappa ^{2}$, We have to look for a model in which a natural 
vacuum expectation value is given to 
$K_{i}F^{i}$ and to give a positive definite mass squared 
to sleptons. In connection with this problem, we should also 
include the sigma-model anomaly. 
Although we did not touch on the sigma-model anomaly 
at all in this paper, we believe that our method is equally 
useful   to analyze the sigma-model anomalous terms. 
All of these  belong to  our future works.

\vspace{1cm}
\noindent
{\Large {\bf Acknowledgement}}

The work of T.K. is supported in part by Grant in Aid 
for Scientific Research from the Ministry of Education 
(grant number 14540265 and 13135215). 



\begin{thebibliography}{99}
%
\bibitem{nilles}
H.P. Nilles, {\it Phys. Reports} {\bf 110} (1984) 1. 
%
\bibitem{giudice}
G.F. Giudice and R. Rattazzi, {\it Phys. Reports } {\bf 322} 
(1999) 25.
%
\bibitem{murayama}
G.F. Giudice, R. Rattazzi, M.A. Luty and H. Murayama, 
{\it JHEP } {\bf 12} (1998) 027. 
%
\bibitem{rs}
L. Randall and R. Sundrum, {\it Nucl. Phys. } {\bf B 557} 
(1999) 79.
%
\bibitem{pomarol}
A. Pomarol and R. Rattazzi, {\it JHEP} {\bf 05} 
(1999) 013.
%
\bibitem{giudice2}
G.F. Giudice and R. Rattazzi, {\it Nucl. Phys.} {\bf B 511} 
(1998) 25.
%
\bibitem{katz}
E. Katz and Y. Shadmi and Y. Shirman, {\it JHEP} {\bf 08} 
(1999) 015.
%
\bibitem{chacko}
Z.  Chacko, M.A. Luty, I Maksymyk and E. Ponton, {\it 
JHEP} {\bf 04} (2000) 001.
\bibitem{allanach}
B.C. Allanach and A. Dedes, {\it JHEP} {\bf 06} (2000) 017.
%
\bibitem{jack}
I. Jack and D.R. Jones, {\it Phys. Letters} {\bf B 482} 
(2000) 167;
\\
M. Carena, K. Huitu and T. Kobayashi, {\it Nucl. Phys. } 
{\bf B 592} (2001) 164;
\\
M. Arkani-Hamed, D.E. Kaplan, H. Murayama and Y. Numura, 
{\it JHEP} {\bf 0102} (2001) 041..
%
\bibitem{jack2}
I. Jack and D.R. Jones, {\it Phys. Letters} {\bf B 491} 
(2000) 151.
%
\bibitem{luty}
M.A. Luty and R. Rattazzi, {\it JHEP} {\bf 9911} (1999) 001.
%
%
\bibitem{moroi}
J.A. Bagger, T. Moroi and E. Poppitz, {\it JHEP} {\bf 0004} (2000) 009.
%
%
\bibitem{waar}
B.J. Waar, Ann. Phys. (N.Y.) {\bf 183} (1988) 1, 59
%
%
\bibitem{kap}
V. Kaplunovsky and J. Louis, {\it Nucl. Phys. } {\bf B 422} (1994) 57;
L.  Dixon, V. Kaplunovsky and J. Louis, {\it Nucl. Phys.} 
{\bf B 355} (1991) 649.
%
\bibitem{ovrut}
G. Lopes Cadoso and B. Ovrut, {\it Nucl. Phys. } {\bf B369} (1992) 351; {\it Nucl. Phys. } {\bf B 392} 
(1993) 315; {\it Nucl. Phys.} {\bf B 418} (1994) 535
%
%
\bibitem{ferrara}
J.P. Deredinger, S. Ferrara, C. Kounnas and F. Zwirner, {\it 
Nucl. Phys. } {\bf B 372} (1992) 145. 
%
%
\bibitem{boyda}
E. Boyda, H. Murayama and A. Pierce, {\it Phys. Rev.}
 {\bf D65} (2002) 085028.
%
\bibitem{fad}
L.D. Faddeev and A.A. Slavnov, {\it Gauge Fields} (The Benjamin Pub. 
Co., 1980) ;
T.D. Bakeyev and A.A. Slavnov, {\it Mod. Phys. Lett.} {\bf A11} 
(1996) 1539.
%
%
\bibitem{wessbagger}
J. Wess and J. Bagger, {\it Supersymmetry and Supergravity} 
(2nd edition, Princeton University Press, 
Princeton, New Jersey, 1992)
%
%
\bibitem{Stepanyantz:2003zb}
K.~V.~Stepanyantz,
arXiv:hep-th/0301167.
\\
A.~A.~Soloshenko and K.~V.~Stepanyantz,
arXiv:hep-th/0304083.
%
%
\bibitem{nishihara}
K. Nishihara and T. Kubota, {\it in preparation}
%
\end{thebibliography}
\end{document}